 \documentclass[final,5p,times]{elsarticle}

\usepackage{amssymb}

\usepackage{lineno}

\usepackage{graphicx}
\usepackage{subcaption}
\usepackage{float}
\usepackage[thinc]{esdiff}
\usepackage{amsmath}
\usepackage{nicefrac}
\usepackage{mathtools}
\usepackage{appendix}
\usepackage{dcolumn}
\usepackage{bm}
\usepackage{widetext}
\usepackage{color,soul}
\usepackage{hyperref}

\begin{document}

\begin{frontmatter}



\title{Asteroid Disruption and Deflection Simulations for Multi-Modal Planetary Defense}


\author[label1]{Alexander N. Cohen}
\ead{ancohen@ucsb.edu}
\author[label1]{Philip Lubin}
\author[label2]{Darrel Robertson}
\author[label3]{Mark Boslough}
\author[label4]{Sasha Egan}
\author[label1]{Brin Bailey}
\author[label5]{Angela M. Stickle}
\author[label6]{Elizabeth A. Silber}
\author[label1]{Peter Meinhold}
\author[label1]{Dharv Patel}

\affiliation[label1]{organization={University of California - Santa Barbara},
            addressline={Broida Hall}, 
            city={Santa Barbara},
            state={California},
            postcode={93106}, 
            country={United States}}
\affiliation[label2]{organization={NASA Ames Research Center},
            city={Moffett Field},
            state={California},
            postcode={94035},             
            country={United States}}
\affiliation[label3]{organization={University of New Mexico},
            addressline={1700 Lomas Blvd, NE. Suite 2200}, 
            city={Albuquerque},
            state={New Mexico},
            postcode={87131},
            country={United States}}
\affiliation[label4]{organization={New Mexico Institute of Mining and Technology, Energetic Materials Research and Testing Center},
            city={Socorro},
            state={New Mexico},
            postcode={87801},
            country={United States}}
\affiliation[label5]{organization={Johns Hopkins University Applied Physics Laboratory},
            city={Laurel},
            state={Maryland},
            postcode={20723},
            country={United States}}
\affiliation[label6]{organization={Sandia National Laboratories},
            city={Albuquerque},
            state={New Mexico},
            postcode={87123},
            country={United States}}

\begin{abstract}
Planetary defense from asteroids via deflective means alone does not offer viable solutions in terminal scenarios where there is little warning time before impact. The PI method of planetary defense enables operation in terminal interdiction modes where there is little warning time prior to impact, but can also operate in the same extended time scale interdiction modes as made possible by traditional deflection techniques, which results in a versatile, multi-modal planetary defense capability. The method is also practical and cost-effective since it relies solely on launch vehicles and penetrator materials already available today, and thus presents itself as a logical and competitive option for planetary defense. As per the PI method, we investigate the effectiveness of rubble pile asteroid disruption and deflection via hypervelocity impacts with 10:1 aspect ratio cylindrical tungsten penetrators. We present the results of an ongoing simulation campaign dedicated to investigating the PI method, using the Lawrence Livermore National Laboratory (LLNL) arbitrary Lagrangian-Eulerian (ALE) hydrodynamics code ALE3D run with the High-End Computing Capability (HECC) at NASA Ames Research Center. We model heterogeneous rubble pile asteroids with a distribution of spherical boulders of varying initial yield strengths set within a weak binder material. We find that rubble pile asteroids of this type in the 20 -- 100 meter-class can be effectively mitigated via 20 km/s impacts with 100 -- 1000 kg penetrators via the coupling of the penetrator kinetic energy into the bulk material of the asteroid. 
\end{abstract}




\end{frontmatter}


\section{\label{sec:intro}Introduction}
PI (``Pulverize It'') is a novel, multi-modal method of planetary defense which can operate in both extended timescale interdiction modes with long warning times and in terminal interdiction modes where there is little warning time. The method can mitigate a bolide threat in a \textcolor{black}{disruptive or deflective manner depending on the size, warning time, and closing speed} \cite{lubin_asteroid_2023,lubin_pi_2023,bailey_optical_2024}. This is achieved via the bolide impacting a high-density, hypervelocity penetrator (or arrays, waves, and various other configurations of penetrators) \textcolor{black}{which, in the case of a disruptive mitigation scenario, disrupts the bolide into acceptably sized fragments. In the case of a deflective scenario, the hypervelocity impact ejects a significant fraction of the target mass which induces greatly enhanced momentum transfer.} 

The hypervelocity impact process rapidly converts a portion of the bolide's kinetic energy into heat and shock waves in the bolide material. Since most bolides arrive at speeds relative to Earth that are faster than those achievable by chemical propulsion ($>10$ km/s), there is only modest gain to be added from the penetrator speed relative to Earth compared to the bolide speed relative to Earth, and therefore the mode of operation is more like placing the necessary configuration of penetrators in the path of the bolide, rather then imparting momentum to it via a rocket-propelled impact. The heat energy of the impact is enough to locally vaporize and ionize material near the penetrator impact sites, and the subsequent shock waves damage and fracture the bolide material as they propagate and refract through it. \textcolor{black}{In the disruptive mode,} this process imparts enough momentum to the bulk of the material to then drive the fragmented bolide apart with enough kinetic energy to overcome the bolide's gravitational binding energy. If necessary, subsequent waves of penetrators in a variety of configurations can enable the reduction of the bolide into acceptably small sized fragments. 

\textcolor{black}{In disruptive interdiction modes with long warning times, the fragment cloud disperses and most, if not all, fragments miss the Earth entirely. In terminal interdiction modes, the fragments enter the Earth's atmosphere and burn up or airburst at high altitude.} Fragments of acceptable size will airburst at altitudes high enough (generally between 30 and 40 km) to eliminate any threat of damage from the generated shock wave, and the dispersion of the fragments both laterally and longitudinally along their path distributes their airburst times and locations significantly. The dispersion of airburst times and locations is key to the success of the method in terminal disruptive modes since it de-correlates the shock wave energy for arbitrary observers on the ground \cite{bailey_optical_2024}. We have shown in our previous work that the acceptable fragment size is a function of the bolide density and exo-atmospheric velocity. For rubble pile asteroids, whose average density is estimated to be between 1 -- 4 g/cm$^3$ \cite{carry_density_2012}, the maximum acceptable fragment size is $\sim10$ meters in diameter \cite{lubin_asteroid_2023,lubin_pi_2023,bailey_optical_2024}.

We begin in Section \ref{sec:ale3d} with a description of the ALE3D code and computational resources used for this study. In Section \ref{sec:penetrators} we describe our high-density, hypervelocity penetrator models, followed by a discussion of our rubble pile asteroid model in Section \ref{sec:rubblepiles}. Section \ref{sec:simparams} discusses the simulation parameters used commonly across all of the simulations presented in this study. In Section \ref{sec:results} and the following subsections, we present simulation results for hypervelocity impacts on 20, 50, and 100 meter diameter rubble pile asteroid targets and compare the differences between impacts with single impactors to those with arrays of impactors of the same total mass. In Section \ref{sec:altmodes} we present alternate modes of operation in which partial fragmentation and high velocity ejecta are used to enhance the momentum transfer in deflective (as opposed to disruptive) scenarios. \textcolor{black}{Finally, in Section \ref{sec:future} we discuss our plans for future work in this ongoing simulation campaign.}

\section{\label{sec:ale3d}ALE3D hypervelocity impact simulations}
In order to investigate the dynamics of hypervelocity impacts in the $>2$ km/s regime, we make use of the Lawrence Livermore National Laboratory (LLNL) arbitrary Lagrangian–Eulerian (ALE) hydrodynamics code ALE3D \cite{noble_ale3d_2017} run with the High-End Computing Capability (HECC) at NASA Ames Research Center. With ALE3D we are able to model hypervelocity impact dynamics in 2D and 3D using equation-of-state material models which include shock response and material vaporization/ionization. We make use of the Livermore Equation of State (LEOS) tables as the building blocks of our material models. The simulation results, which will be discussed in the following sections, are used to inform the design of more efficient penetrators and to better understand the effect of material properties on the dynamics of hypervelocity impact events and total disruption of the target. The results so far generate confidence that \textcolor{black}{robust disruption and enhanced deflection} across the range of plausible asteroid sizes and types can be achieved using minimum impactor mass.

\subsection{\label{sec:penetrators}Penetrator material model}
A 10:1 aspect ratio cylindrical tungsten penetrator with density $19.24$ g/cm$^3$ is used as the baseline projectile for this study, the mass and multiplicity of which is varied across different simulations. \textcolor{black}{An idealized 100 kg version of this penetrator has a height of $h_p\approx87.4$ cm and radius $r_p\approx8.7$ cm, and a 500 kg version has a height of $h_p\approx150.0$ cm and radius $r_p\approx15.0$ cm. However, the actual penetrator mass used in a given simulation varies from the user-specified value and more closely approximates it with increasing mesh resolution. For the simulations presented here, we use mesh resolutions for which the measured penetrator mass differs from the idealized value by less than a factor of two. For all calculations in the following sections (unless specified as idealized values), the measured value of the penetrator mass after initialization $m_p$ is used rather than the user-specified value.} We make use of LEOS table 740 for tungsten in combination with a bilinear yield curve model with 750 MPa yield stress to describe the tungsten penetrator material. 

\begin{figure}
    \centering
    \includegraphics[width=0.48\textwidth]{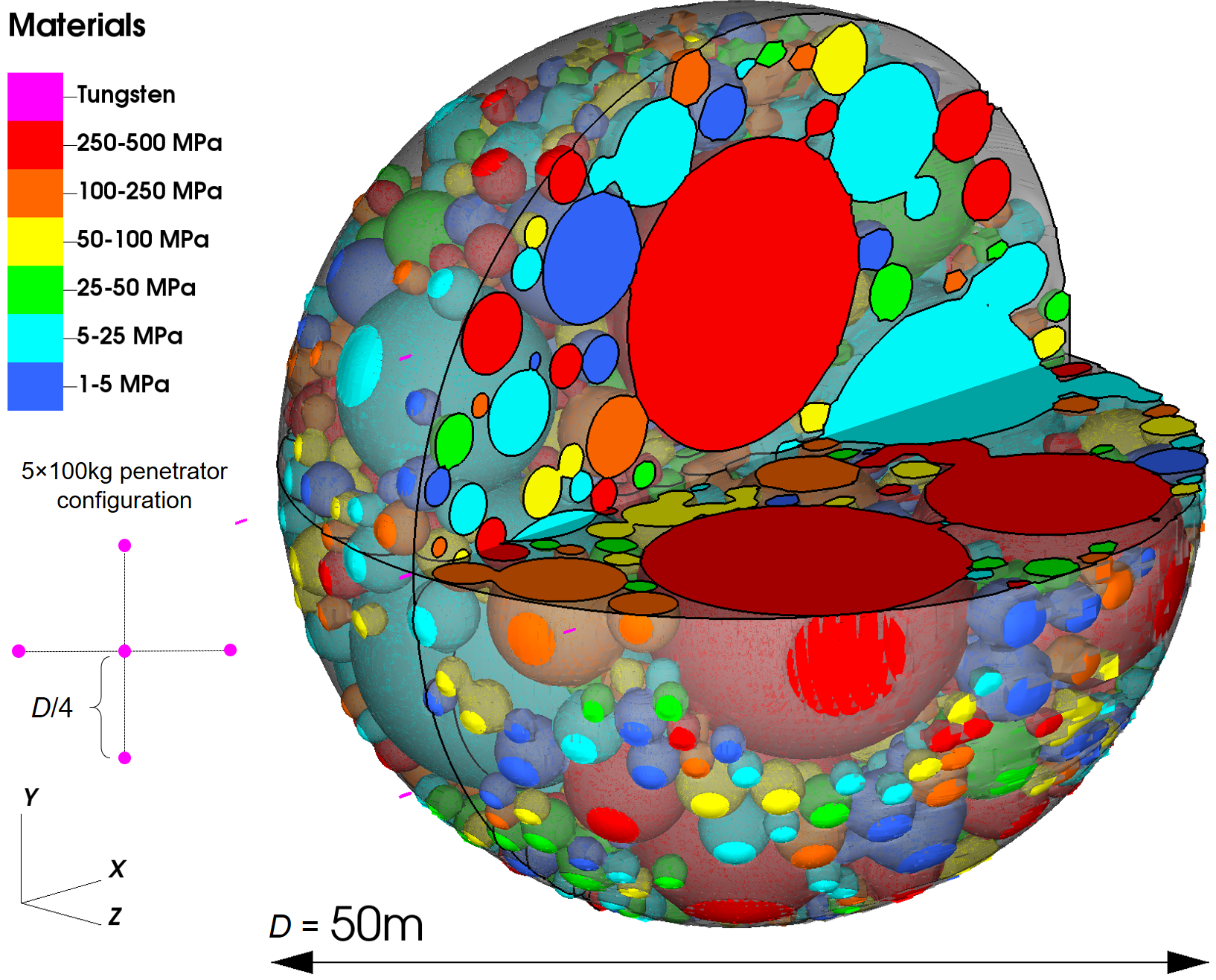}
    \caption{Example rubble pile asteroid model shown in partial cross section for a 50 meter diameter bolide. The binder material is shown in transparent grey, and the boulder distribution within it is colored by the material yield strength ranges, as in Figure \ref{fig:weibull}. An example ``+" configuration of five 100 kg 10:1 aspect ratio cylindrical tungsten penetrators (referred to henceforth as a $5\times100$ kg case) is shown as an inset image on the left, where the penetrators are spaced by a distance equal to \textcolor{black}{half the radius} of the bolide.}
    \label{fig:materials}
\end{figure}

\subsection{\label{sec:rubblepiles}Rubble pile asteroid model}
Our baseline bolide model consists of a sphere of weak binder material and a distribution of spherical boulders embedded within it, as illustrated in Figure \ref{fig:materials}. Depending on the particular bolide model, the sizes and packing densities of the boulders within the distribution are determined by specifying discrete volume fill fractions. The boulders are spatially distributed randomly so as to achieve the desired fill fraction. 

\begin{figure}
    \centering
    \includegraphics[width=0.48\textwidth]{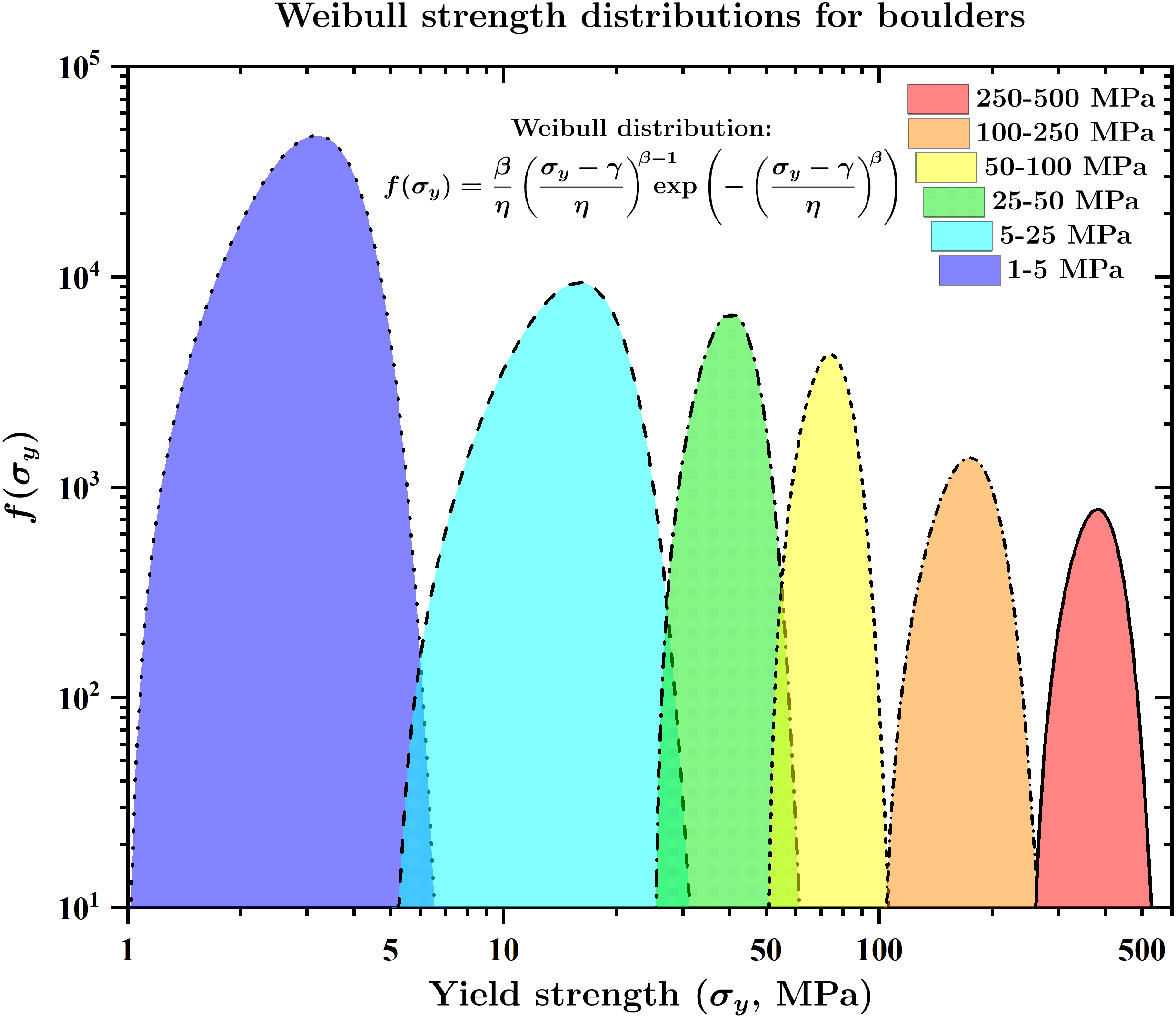}
    \caption{Weibull strength distributions for the six boulder types in our rubble pile asteroid models. These distributions are used to initialize the yield strengths of the boulders, scaling up from 1 MPa initial yield strength. In addition to these, a Weibull distribution is used for the binder material with a weak mean strength of 25 Pa. For reference, the violet 1--5 MPa distribution is comparable to hardened soil, the cyan 5--25 MPa distribution to standard grade concrete, the green 25--50 MPa distribution to high strength concrete, the yellow 50--100 MPa distribution to aluminum, the orange 100--250 MPa distribution to structural steel, and the red 250--500 MPa distribution to high strength steel and titanium. These strengths are an extremely conservative over-estimation of the strength of rubble pile asteroids \cite{sanchez_strength_2014,pravec_fast_2000}.}
    \label{fig:weibull}
\end{figure}

For our rubble pile asteroid models, we use LEOS table 4030 for granite with density 2.67 g/cm$^3$ as the baseline material model for both the weak binder material and the boulder distribution. \textcolor{black}{Granite is chosen as a general silicate-rich material (primarily composed of quartz and feldspar) with a conservatively high strength, though as will be discussed we vary the strength of this material model significantly. See Section \ref{sec:future} for a discussion on future work with more diverse material models.} The boulder distribution is divided into six boulder types which differ from each other in their average compressive yield strength, as illustrated in Figure \ref{fig:weibull}. Each boulder type is initialized with a Weibull distribution of strengths given by 
\textcolor{black}{\begin{equation}
    f(\sigma_y)=\mathop{\frac{\beta}{\eta}}\bigg(\frac{\sigma_y-\gamma}{\eta}\bigg)^{\beta-1}\exp{\left(-\bigg(\frac{\sigma_y-\gamma}{\eta}\bigg)^\beta\right)},
\end{equation}
where $\sigma_y$ is the material yield strength and} the three parameters $\beta$, $\eta$, and $\gamma$ control the distribution shape, scale, and location, respectively, which enables the modeling of fracture dynamics in heterogeneous rocky materials. Similarly, the binder material is initialized with a Weibull distribution of yield strengths with a weak average strength of 25 Pa, as per the results of a study by S\'anchez and Scheeres which suggests that the cohesive strength in rubble piles asteroids may be quite weak \cite{sanchez_strength_2014}. 

\begin{figure}
    \centering
    \includegraphics[width=0.48\textwidth]{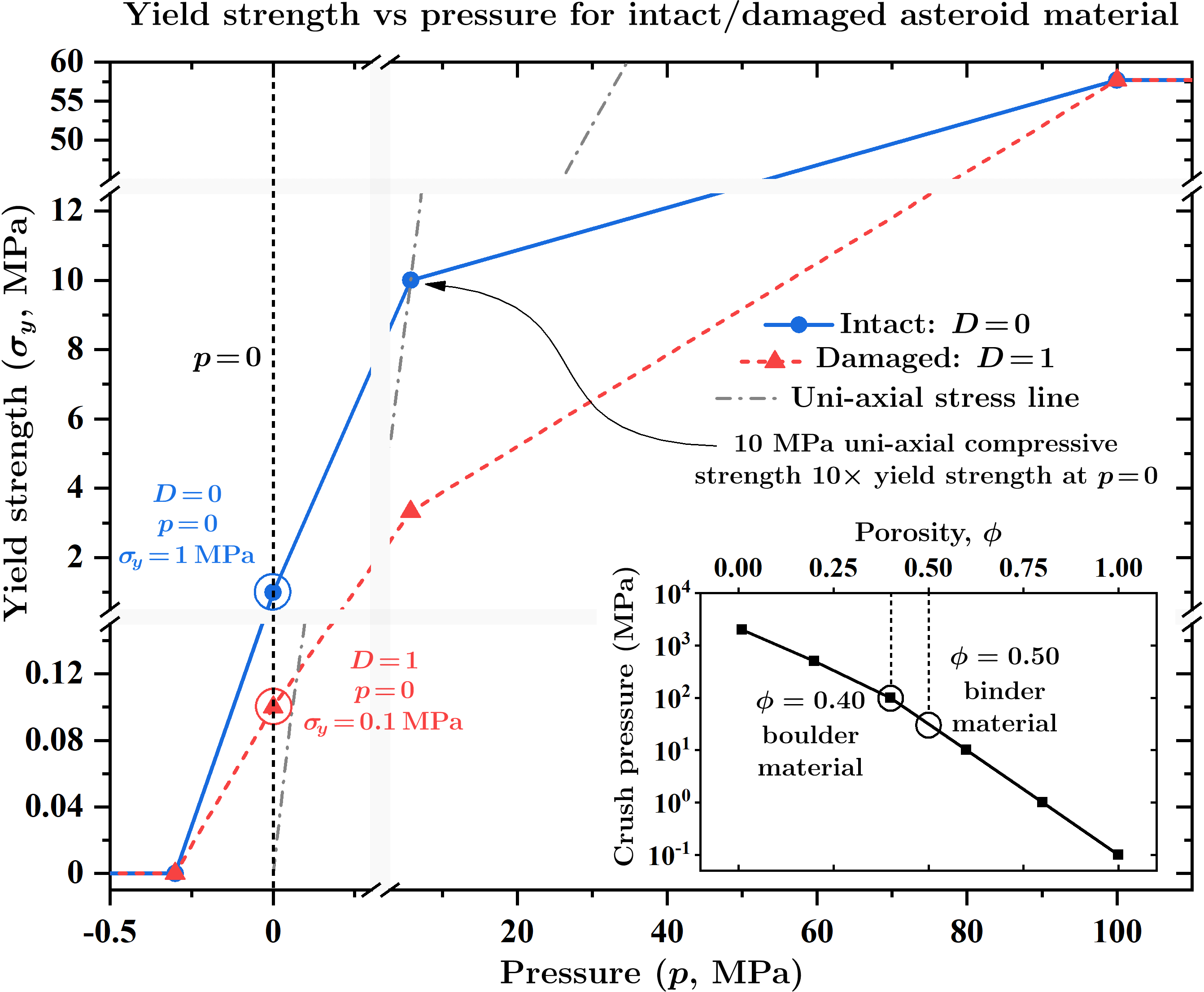}
    \caption{Compressive yield strength $\sigma_y$ vs applied pressure $p$ comparison between \textcolor{black}{damaged ($D=1$) and intact ($D=0$)} asteroid material with 1 MPa initial yield strength at $p=0$. Note that these curves scale with the strength of each particular boulder as in Figure \ref{fig:weibull}. The uni-axial stress line $\sigma_y=\sqrt{3}\mathop{k_i}$ is shown as a grey dot-dash line, where $k_i$ is the uni-axial stress in the direction $\hat{i}$. The intersection of the uni-axial stress line with the intact material yield strength curve (solid blue line with circular symbols) indicates the material's uni-axial compressive strength, 10 MPa in this case, which is $10\times$ the material yield strength at zero pressure. \textbf{Inset:} Crush pressure in MPa vs material porosity, as per \cite{housen_impacts_2018}. In our rubble pile asteroid models, the boulders are initialized with 40\% porosity and the binder material with 50\% porosity, corresponding to crush pressures of 100 and $\sim30$ MPa, respectively.}
    \label{fig:yieldvspressure}
\end{figure}

We also add a porous crush model to the binder and boulder materials, \textcolor{black}{which modifies the material density $\rho_s$ and equation of state (EOS) to model the effect of porosity in the presence of shocks. The effective density of porous material is given by
\begin{equation}
    \rho=(1-\phi)\mathop{\rho_s},
\end{equation}}
where $0\leq\phi<1$ is the porosity. The pressure in the material is therefore given by
\begin{equation}
    p = (1 - \phi)\mathop{p_s},
\end{equation}
where $p_s(\rho_s)$ is the pressure-density relationship of the material in solid phase, \textcolor{black}{supplied in this case by the LEOS 4030 tabular equation of state.} Effectively, as pressure is applied the porosity must be crushed out by a pressure $p$ exceeding the crush pressure $p_c$ before any damage to the material takes place. We specify the initial crush pressure as 100 MPa \textcolor{black}{(40\% porosity)} for the boulder material and $\sim30$ MPa \textcolor{black}{(50\% porosity)} for the binder material, as shown in the inset of Figure \ref{fig:yieldvspressure}, as per a study investigating the effect of porosity on hypervelocity impact crater formation \cite{housen_impacts_2018}. 

Damaged material is assumed to no longer support negative pressures, but still retains compressive strength. Therefore, the EOS properties of damaged and intact material differ only slightly, as shown in Figure \ref{fig:yieldvspressure}, except for when $p<0$, for which the pressure in damaged material is modified by
\begin{equation}
    p_d=\mathop{(1-D)}p_i,
\end{equation}
where $p_i$ is the equivalent pressure for intact material and $0\leq D\leq 1$ is the material damage parameter.

\begin{figure*}
    \centering
    \includegraphics[width=0.99\textwidth]{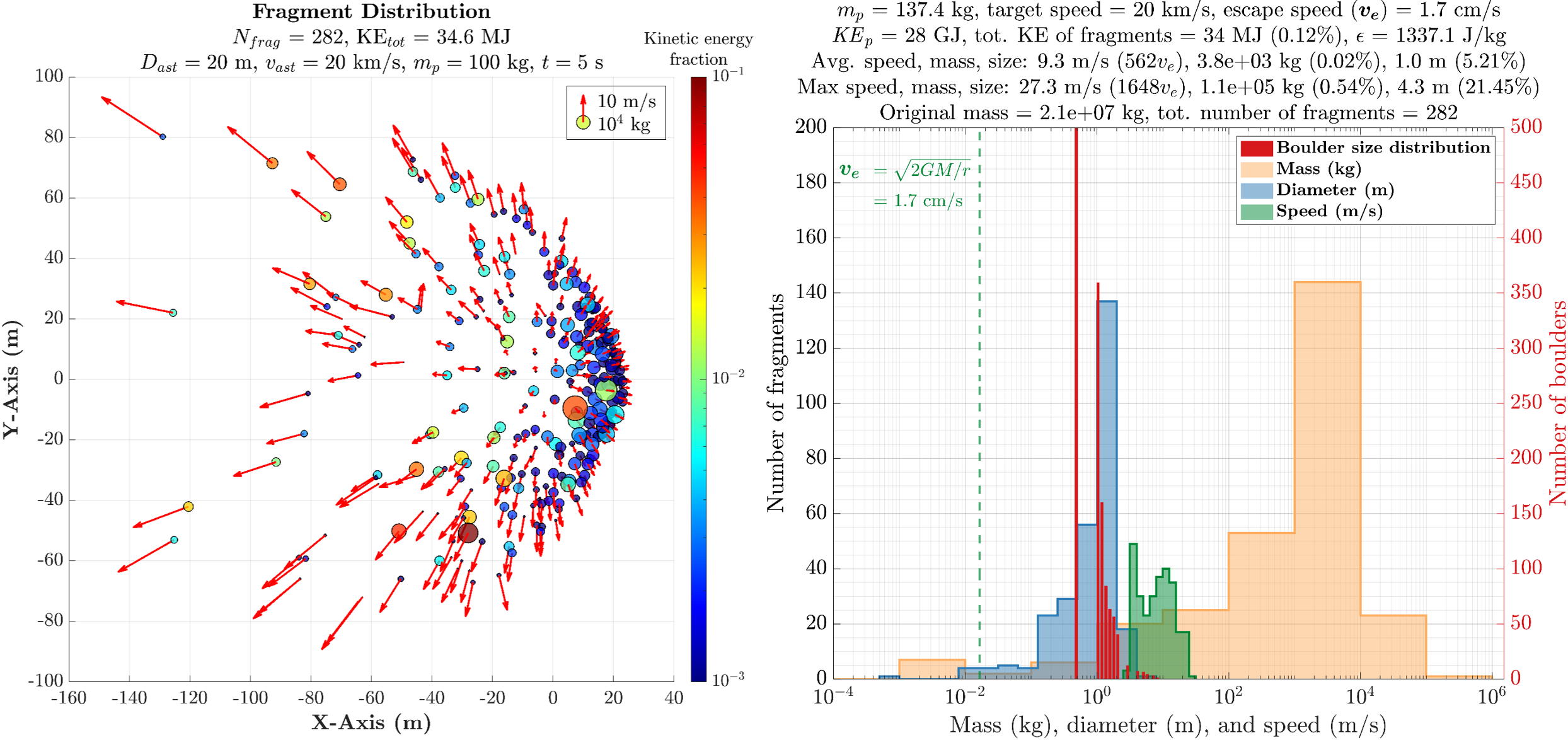}
    \caption{\textcolor{black}{Fragment distribution and statistics at $t=5$ seconds after impact for a 137.4 kg 10:1 aspect ratio tungsten cylindrical penetrator incident at 20 km/s upon a 20 m diameter rubble pile asteroid target. The orange, blue, and green histograms indicate the distributions of fragment masses in kilograms, diameters in meters, and speeds in meters per second, respectively. The red histogram (corresponding to the right $y$-axis) shows the original boulder size distribution. Of note is the average fragment size of 1 m and average fragment speed of 9.3 m/s, which is over $500\times$ greater than the gravitational escape speed of $v_e=1.7$ cm/s (dashed green line). Also note that the maximum fragment size is 4.3 m, which is well below the 10 m acceptable fragment size threshold for rocky asteroid densities around 2.6 g/cm$^3$ \cite{lubin_asteroid_2023,lubin_pi_2023}. These results suggests that Chelyabinsk-like asteroids can be mitigated using a single tungsten penetrator with mass on the order of 100 kg, assuming a bolide closing speed of 20 km/s. With an original parent bolide mass of $2.1\times10^7$ kg and 28 GJ of penetrator kinetic energy, we have a specific impact energy of 1333 J/kg, which corresponds to the conclusion that catastrophic disruption should be achieved by greatly exceeding the 100 J/kg limit \cite{jutzi_fragment_2010}.}}
    \label{fig:20m}
\end{figure*}

\subsection{\label{sec:simparams}Simulation parameters}
We simulate a hypervelocity impact event in two phases: an early-time phase with high spatial and temporal resolution, which is then followed by a lower resolution phase which enlarges the simulated region and extends the simulation to macroscopic timescales. In the first phase, which typically ranges from $t=0$ to 10 milliseconds after impact, we resolve the system with a static 3D Eulerian mesh to a size scale at least half of the penetrator radius near each penetrator impact site, with the mesh density decreasing with distance from each impactor site. The second phase of the simulation typically ranges from $t=10$ milliseconds after impact out to $\geq10$ seconds, and uses a static 3D Eulerian mesh with a constant size of 1 m$^3$ throughout the simulated region. 

Low density asteroid and penetrator material is important for modeling the early time dynamics of the first phase, as will be described further in Section \ref{sec:altmodes}, but it becomes a computational burden in the second phase of the simulation where low density material is no longer actively helping to drive the disruption of the bolide. In order to decrease computational complexity in the second phase of the simulation, \textcolor{black}{and because the low density material is no longer a dominant contributor to the disruption of the bolide,} we introduce a threshold criterion by which low density asteroid and penetrator material below $10^{-3}$ g/cm$^3$ can be removed from the calculation. This enables the extension of the simulation to macroscopic timescales. \textcolor{black}{No significant difference in disruptive behavior has been observed between simulations with and without the low density material removal threshold in place.}

Figure \ref{fig:materials} shows an example of a 50 meter diameter rubble pile asteroid at the start of an early time simulation with five 100 kg penetrators. The penetrators start with 20 km/s speed relative to the asteroid rest frame and are oriented such that their length axes and velocity vectors are both parallel to the $z$-axis. The origin of the simulation is located at the center of the asteroid sphere.

\begin{figure}
    \centering
    \includegraphics[width=0.48\textwidth]{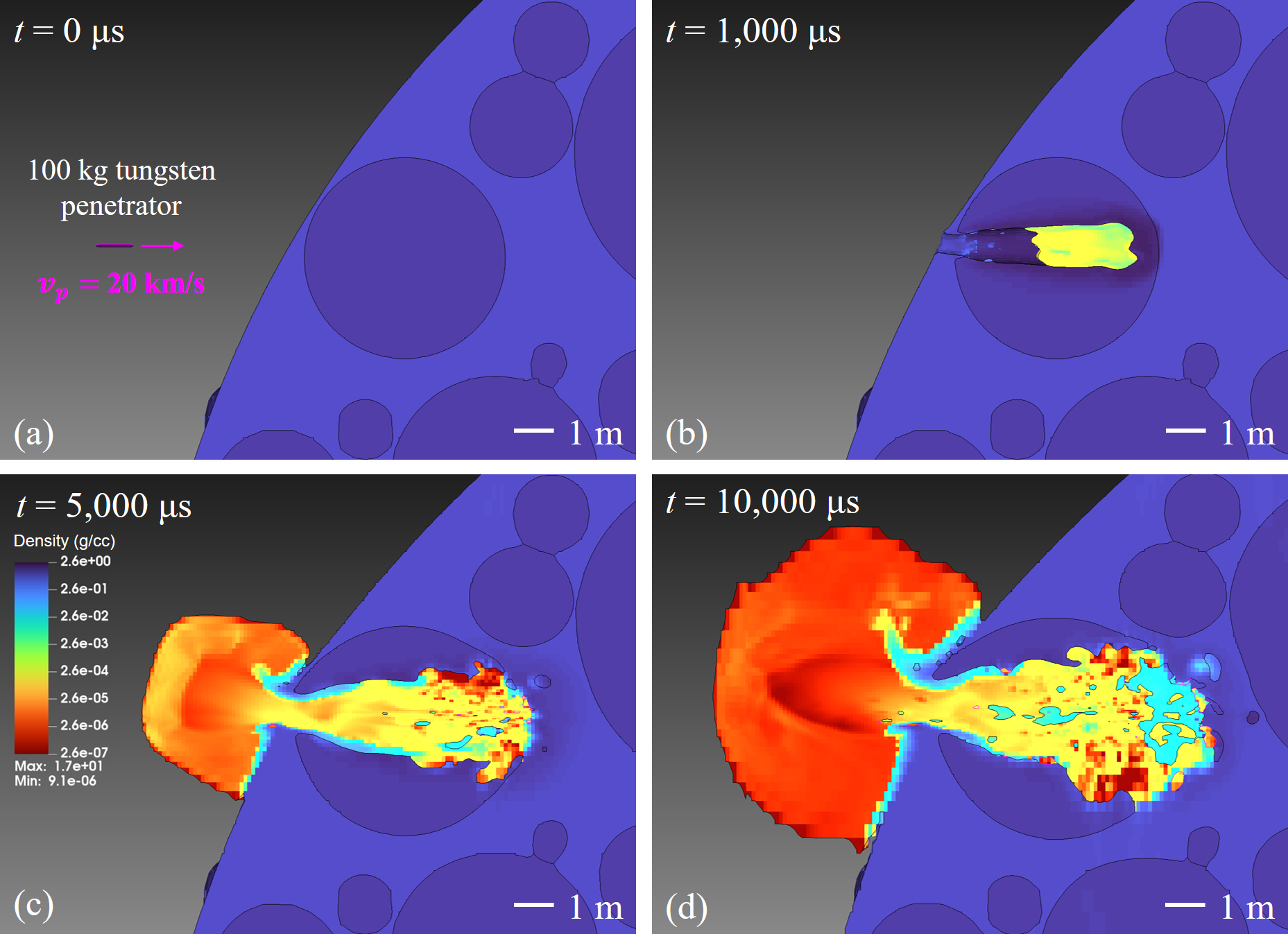}
    \caption{Sequence of early time frames showing a single 100 kg 10:1 aspect ratio cylindrical tungsten penetrator incident at 20 km/s upon a 50 meter rubble pile asteroid target. The color scale is a logarithmic function of the material density ranging from 2.6 to $2.6\times10^{-7}$ g/cm$^3$. Frame (a) shows the simulation in its initial state, with the penetrator highlighted in magenta. Frame (b) shows the simulation 1,000 microseconds after impact. By this time, the penetrator has been completely vaporized and is now mixed with local asteroid material which has been vaporized and heated to temperatures well in excess of 5000 K. Frame (c) shows the simulation 5,000 microseconds after impact. The vaporized material previously contained within the entrance cavity is ejected at speeds exceeding 1 km/s. Frame (d) shows the simulation 10,000 microseconds after impact, where a penetration depth $>10$ m is achieved.}
    \label{fig:earlytime}
\end{figure}

\begin{figure*}
    \centering
    \includegraphics[width=0.99\textwidth]{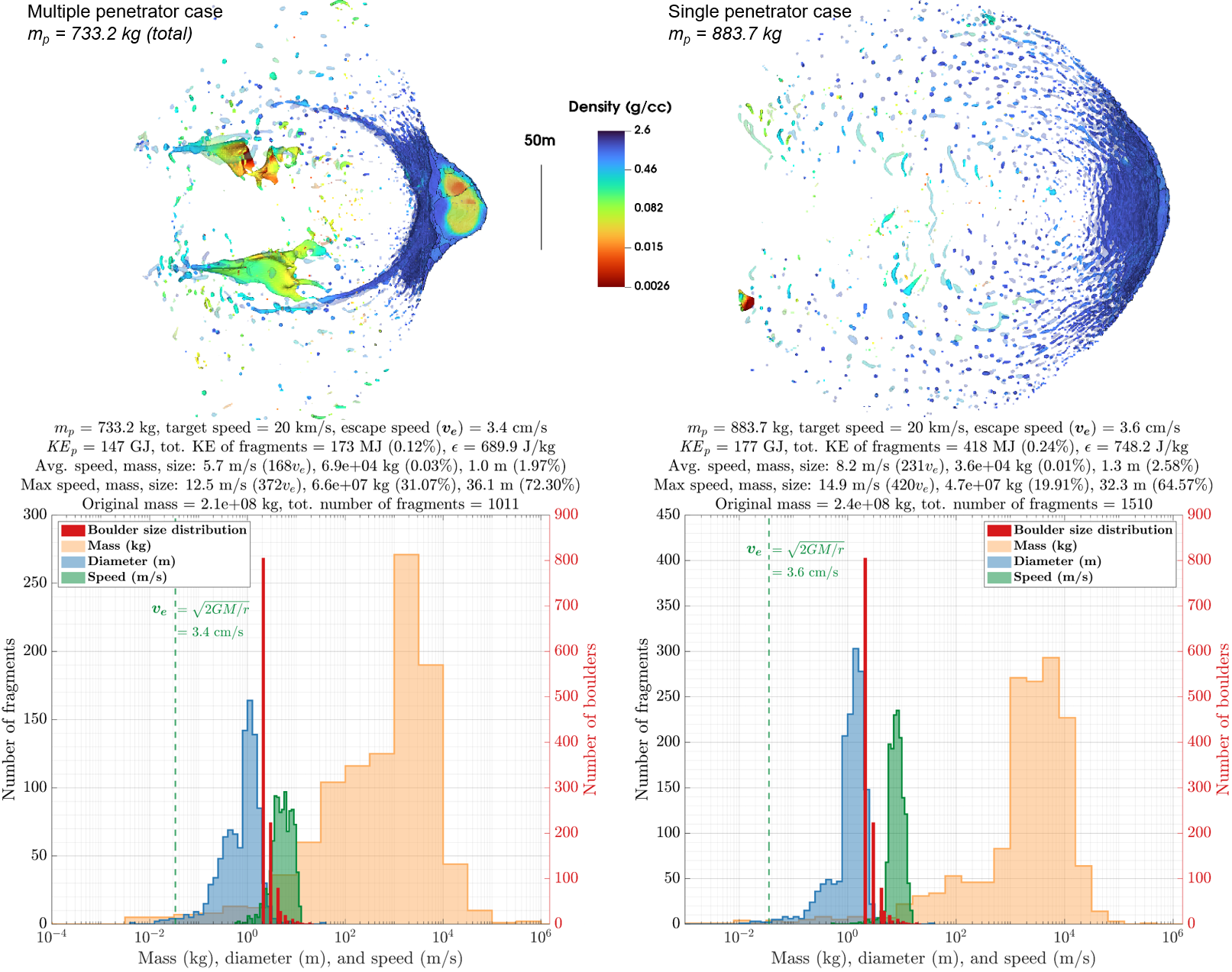}
    \caption{\textcolor{black}{Fragment statistics at $t=10$ seconds after impact for the multiple penetrator case (left) and the single penetrator case (right), both incident at 20 km/s upon identical 50 m diameter rubble pile asteroid targets. The orange, blue, and green histograms indicate the distributions of fragment masses in kilograms, diameters in meters, and speeds in meters per second, respectively. The red histogram (corresponding to the right $y$-axis) shows the original boulder size distribution. When compared to the blue histogram of fragment sizes 10 seconds after impact, it can be seen how the initial boulders have been largely reduced into fragments with $\sim1$ m average diameter in both cases. Also shown (top) is a visualization of each case at $t=10$ seconds after impact, with the color scale corresponding to material density in g/cm$^3$. Note that the large regions of low density ($< 10^{-2}$ g/cm$^3$) material are not included in the fragment count. The effectiveness of disruption can be quantified by comparing the 100 GJ initial kinetic energy of the penetrator(s) to the total kinetic energy of the fragments at a later time. In these cases, we see 0.12\% of the original kinetic energy has been transferred to the fragments in the multiple penetrator case, and 0.24\% has been transferred to the fragments in the single penetrator case. It is clear from these results that the single penetrator case succeeds in transferring $\sim2\times$ more of the initial kinetic energy of the penetrator than the multiple penetrator case. The rest of the initial penetrator kinetic energy is lost to thermal effects and to unresolved material below the 1 m$^3$ mesh resolution of the simulation. There are a total of 1,011 fragments at $t=10$ seconds after impact in the multiple penetrator case, and 1,510 fragments in the single penetrator case, but at this time both cases still exhibit a larger, more persistent fragment at the right edge. In both cases, this larger fragment is in the process of disassembling on timescales longer than 10 seconds as evidenced by its rapid and continued deformation due to being comprised of mostly completely failed material. Material in the failed state is only bound by frictional cohesion and gravitational binding, which is greatly overcome by the fragment speeds induced by the impact. These reasons in combination lend credence to the hypothesis that the largest remaining fragment will dissociate into fragments of size comparable to the original interior boulder distribution, though likely even smaller due to the boulder material having been failed.}}
    \label{fig:5x100kgvs1x500kg}
\end{figure*}

\section{\label{sec:results}Results}
For the purposes of this paper, we have simulated impact events with rubble pile bolide targets ranging from diameter 20 -- 100 meters. For 20 km/s impact events, 100 and 500 kg \textcolor{black}{(idealized values)} penetrators arrive with kinetic energies in the reference frame of the bolide of 20 and 100 GJ, respectively.  \textcolor{black}{If we define 
\begin{equation}
    \Omega = {E_p}/{E_\textrm{BE}}
\end{equation}
as the ratio between the penetrator kinetic energy $E_\text{p}$ and gravitational binding energy $E_\text{BE}$}, as in Table \ref{tab:ratios}, it can be seen how even an extremely modest coupling of the penetrator kinetic energy to the bulk material of the bolide can result in disruption beyond the possibility of gravitational recombination. 
\begin{table}[h]
    \centering
    \begin{tabular}{c|c|c|c}
         Diam. (m) & 20 & 50 & 100 \\
         \hline $E_\text{BE}$ (kJ) & 0.12 & 11.57 & 370.30 \\
         \hline $\Omega_{100kg}$ & $1.7\times10^8$ & $1.7\times10^6$ & $5.4\times10^4$ \\
         \hline $\Omega_{500kg}$ & $8.3\times10^8$ & $8.6\times10^6$ & $2.7\times10^5$ \\
         \hline $\epsilon_{100kg}$ (J/kg) & 3616.0 & 235.9 & 29.4 \\
         \hline $\epsilon_{500kg}$ (J/kg) & - & 1180.5 & 148.0 \\
    \end{tabular}
    \caption{Gravitational binding energies $E_\text{BE}$ for 20, 50, and 100 m diameter bolides. Also shown are the ratios $\Omega$ between the gravitational binding energy of the bolide and the penetrator kinetic energies for 100 and 500 kg (idealized values) penetrators at 20 km/s, the large order of magnitude of which suggest that even modest coupling of penetrator kinetic energy to the bulk material of the bolide should result in disruption. The specific impact energies for 100 and 500 kg penetrators are also shown, \textcolor{black}{which are useful to compare to the disruption limit of 100 J/kg found in} \cite{jutzi_fragment_2010}. Note that the specific impact energy for the 500 kg case is not shown for the 20 m bolide since catastrophic disruption is achieved by the 100 kg case, as shown in Figure \ref{fig:20m}.}
    \label{tab:ratios}
\end{table}
Fitted simulation data suggest a limit on the disruption energy per unit target mass (from here on referred to as the specific impact energy) of about $\epsilon=100$ J/kg for sphere-on-sphere type impacts, above which catastrophic disruption is achieved, which they define as having at least 50\% of the original parent bolide mass having been permanently removed \cite{jutzi_fragment_2010,jutzi_sph_2015}. \textcolor{black}{This value is a lower limit and lies below any variations induced by target size, porosity, and friction (see Figure 5 in \cite{jutzi_sph_2015}). While the case is different for specifically designed penetrators which can more efficiently couple their kinetic energy to the bolide, it is a still a useful comparison.} 

\subsection{\label{impacts}Hypervelocity penetrator impact dynamics}
For 100 kg 10:1 aspect ratio tungsten penetrators, asteroid material local to the impact site is rapidly vaporized and ionized up to $\sim10$ m below the surface, which creates a well-tamped explosively expanding volume consisting of the high temperature vaporized asteroid and penetrator material. 500 kg penetrators increase this penetration depth to $\sim15$ m. This tamped hot gas explosion then acts as a gas engine over longer time scales, the high internal pressure of which drives material back out of the impact site at high speeds (like a rocket engine) and also drives the rest of the local non-vaporized asteroid material radially outwards (see the sequence of frames in Figure \ref{fig:earlytime}). This behaviour suggests alternative modes of operation in addition to the extended timescale and terminal interdiction modes, as will be discussed in Section \ref{sec:altmodes}.

\subsection{\label{20m}20 meter-class bolides (robust disruption)}
For 20 meter-class bolides of the rubble pile type described above \textcolor{black}{(similar in size to the Chelyabinsk asteroid from 2013, though not necessarily in internal composition)}, simulation results suggest that a single 100 kg 10:1 aspect ratio tungsten cylinder penetrator impacting at 20 km/s is sufficient for disruption of the bolide down to acceptably small fragment sizes. As an example, Figure \ref{fig:20m} shows the fragment distribution $t=5$ seconds after impact of a penetrator of the type described above \textcolor{black}{(measured penetrator mass $m_p=137.4$ kg)}. Of the 282 fragments resolved above 1 m$^3$ in the simulation, the largest fragment has a diameter of 4.3 meters, well below the acceptable threshold of $\sim10$ meters for fragments of density 2.6 g/cm$^3$ \cite{lubin_asteroid_2023,lubin_pi_2023}. \textcolor{black}{The average fragment speed is 9.3 m/s, which is over $500\times$ the parent bolide's gravitational escape speed of $v_e=1.7$ cm/s, and the kinetic energy of all 282 fragments amounts to 34 MJ, or approximately 0.12\% of the original 28 GJ of kinetic energy delivered by the penetrator. This yields a specific impact energy of $\epsilon\approx1337$ J/kg, which is well above the 100 J/kg disruption limit found in \cite{jutzi_fragment_2010}, and as expected, we indeed catastrophically disrupt the bolide.}  

\subsection{\label{50m}50 meter-class bolides (robust disruption)}
As the size of the target bolide increases, so does the necessary total mass in penetrators to achieve disruption. \textcolor{black}{By increasing the target bolide diameter from 20 to 50 m, we find we can achieve sufficient disruption by also scaling the total penetrator mass to $\sim1000$ kg (idealized).} In order to compare the effect of distributing the penetrator mass in an array of smaller penetrators, as opposed to concentrating it into a single penetrator, we simulate the following two cases: \textcolor{black}{a single $\sim884$ kg penetrator case and a multiple penetrator case with five $\sim147$ kg penetrators arranged in a ``+" shape (total penetrator mass $\sim733.2$)}, as illustrated in Figure \ref{fig:materials}. \textcolor{black}{Figure \ref{fig:5x100kgvs1x500kg} shows simulation results and fragment statistics at $t=10$ seconds after impact for the single and multiple penetrator cases. The specific impact energies for the single and multiple penetrator cases are $\sim748$ J/kg and $\sim690$ J/kg, respectively. Thus, we should expect catastrophic disruption in both cases. Of the 177 GJ of kinetic energy delivered in the single penetrator case, approximately 418 MJ, or 0.24\%, is transferred to the fragments, while only 173 MJ, or 0.12\%, of the 147 GJ delivered in the multiple penetrator case is transferred to the fragments. This results in average fragment speeds of 8.2 m/s and 5.7 m/s for the single and multiple penetrator cases, respectively, both of which are greater than $100\times$ the gravitational escape speeds of the original parent bolides. Of interest is the difference in transfer of kinetic energy to the fragments in both cases, where we see the single penetrator case succeeding in transferring $\sim2\times$ more of the initial kinetic energy of the penetrator than the multiple penetrator case. The early time simulations show that the concentration of the penetrator mass into a single impactor results in greater depth of penetration, and thus greater coupling of the penetrator kinetic energy to the bulk kinetic energy of the fragments due to better tamping of the hot gas explosion produced by the impact.} 

\subsection{\label{100m}100 meter-class bolides (enhanced deflection)}
\begin{figure}
    \centering
    \includegraphics[width=0.48\textwidth]{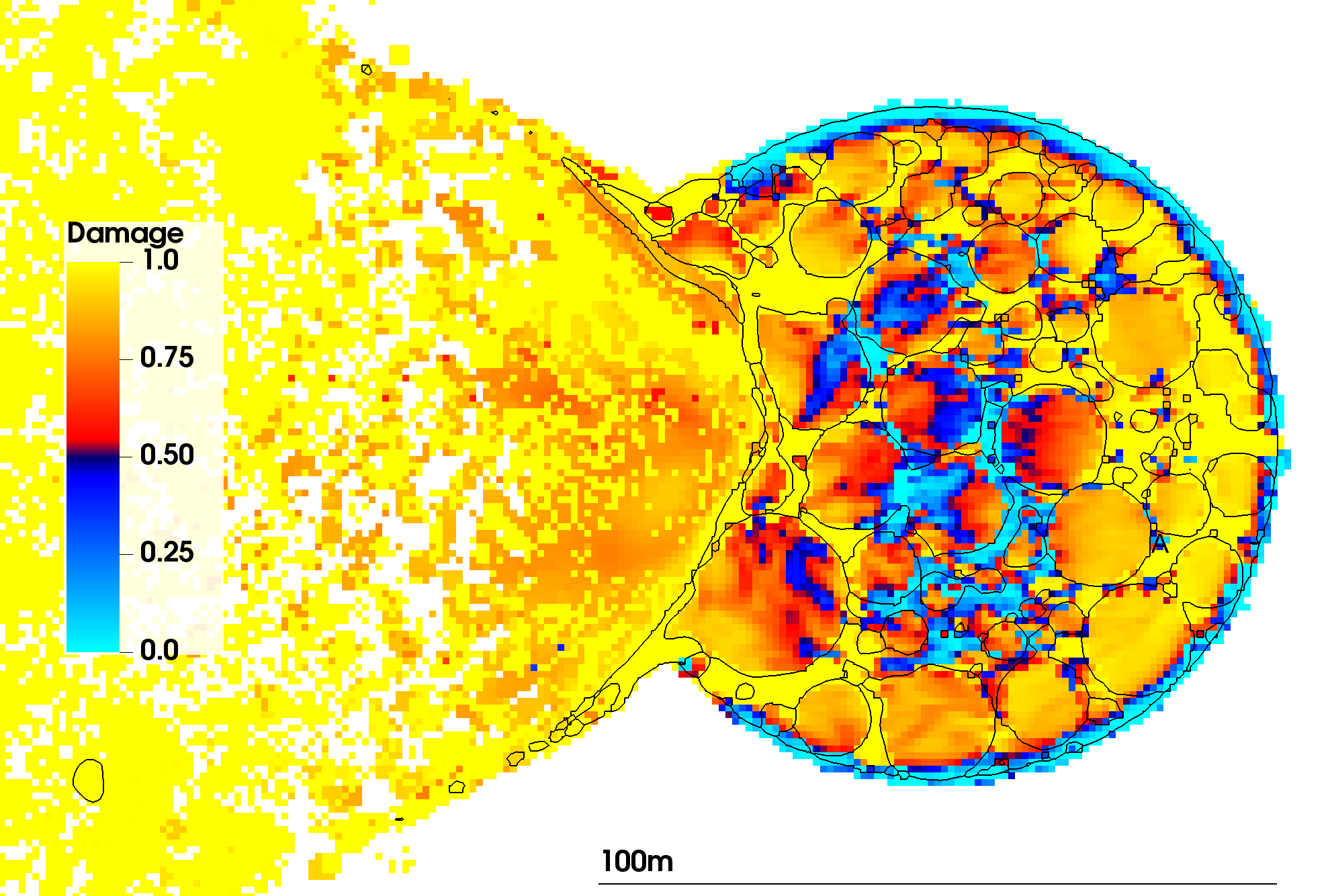}
    \caption{\textcolor{black}{Material damage parameter $0\leq D\leq1$ plot at $t=10$ seconds after impact of a 164.6 kg 10:1 aspect ratio tungsten cylindrical penetrator incident at 20 km/s upon a 100 m diameter rubble pile asteroid target.} Note that the plot shows an equatorial cross section of the spherical bolide, and that the color scale is such that warm colors indicate $D\geq0.5$ and cool colors indicate $D\leq0.5$. At this time the damage parameter takes an average value of \textcolor{black}{$\langle D\rangle=0.76$.} The results of this simulation agree with the prediction that a specific impact energy of $\epsilon_\text{100kg}=29.4$ J/kg should not result in catastrophic disruption of the bolide, and thus suggest that greater penetrator mass is required (see Figure \ref{fig:100m5x100kgDamage}). \textcolor{black}{However, we measure $\Delta\hspace{-0.5mm} V \approx 10.0$ cm/s, which for an original target mass of $M\approx1.9\times10^9$ kg results in a momentum enhancement factor of $\beta>50$. This is an example of the enhanced deflection mode made possible by the PI method.}}
    \label{fig:100m100kgDamage}
\end{figure}

As can be seen in Table \ref{tab:ratios}, for a single 100 kg penetrator incident upon a 100 m target bolide, the specific impact energy $\epsilon_\text{100kg}=29.4$ J/kg falls well below the 100 J/kg catastrophic disruption \textcolor{black}{limit found} in \cite{jutzi_fragment_2010}. This suggests that the penetrator does not bring enough kinetic energy per unit mass of the target to sufficiently disrupt the bolide, if the specific impact energy dependence for sphere-on-sphere impacts is to be comparable to that for penetrator-on-sphere impacts as simulated here. \textcolor{black}{In Figure \ref{fig:100m100kgDamage} we show the results of a simulation of an impact between a 20 km/s penetrator of mass $m_p=164.6$ kg and a 100 m diameter target, where we plot the material damage parameter $0\leq D\leq1$ at a time $t=10$ seconds after impact.} At this time, the damage parameter takes on an average value of \textcolor{black}{$\langle D\rangle=0.76$.} After this time, the shock waves of speed 2.2 -- 2.5 km/s (a function of the variably distributed asteroid porosity and material strengths) produced by the impact have long since dissipated and the bolide continues to deform inertially. This leaves the undamaged regions to persist in their dimensions and potentially allows regions of adjacent undamaged material to remain cohesively connected despite having dramatically weakened the average cohesive strength of the bolide as a whole. 

\textcolor{black}{Of interest is the momentum enhancement factor $\beta$ achieved by the impact, which we define as the ratio between the momentum change induced by the impact in the target and the momentum of the penetrator prior to impact:
\begin{equation}
    \beta = \frac{M\mathop{\Delta\hspace{-0.5mm} V}}{m_p\mathop{v}},
\end{equation}
where $M$ and $m_p$ are the target and penetrator masses, $\Delta\hspace{-0.5mm} V$ is the change in velocity of the target, and $v=20$ km/s is the impact speed \cite{holsapple_momentum_2012,holsapple_point_1987,housen_deflecting_2012}. At $t=10$ seconds after impact for the case described above with penetrator mass $m_p=164.6$ kg, we measure $\Delta\hspace{-0.5mm} V \approx 10.0$ cm/s, which for an original target mass of $M\approx1.9\times10^9$ kg results in a momentum enhancement factor of $\beta>50$. Future work will extend the timescale of such simulations, which will likely further improve the momentum enhancement factor as more fragments are shed from the expanding crater rim.}
 
\begin{figure}
    \centering
    \includegraphics[width=0.45\textwidth]{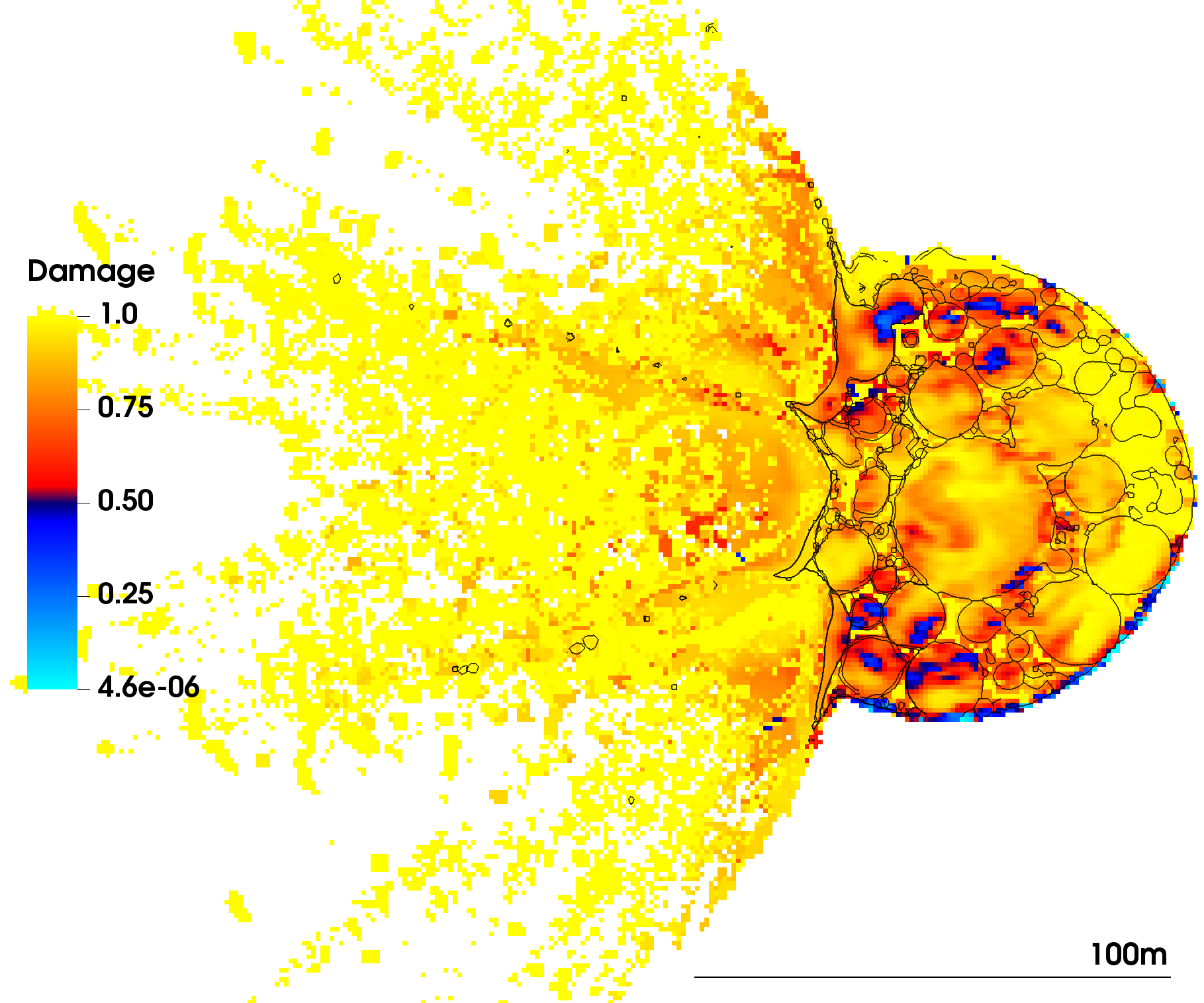}
    \caption{\textcolor{black}{Material damage parameter $0\leq D\leq1$ plot at $t=10$ seconds after impact of a five-penetrator array (arranged in a similar ``+" pattern as in the 50 m case shown previously) of 10:1 aspect ratio tungsten cylindrical penetrators with total mass 1264 kg incident at 20 km/s upon a 100 m diameter rubble pile asteroid target.} As in Figure \ref{fig:100m100kgDamage}, note that the plot shows an equatorial cross section of the spherical bolide, and that the color scale is such that warm colors indicate $D\geq0.5$ and cool colors indicate $D\leq0.5$. \textcolor{black}{At this time the damage parameter takes an average value of $\langle D\rangle=0.80$, or approximately 14.1\% greater than the 164.6 kg single penetrator case shown in Figure \ref{fig:100m100kgDamage}. At this time, we measure $\Delta\hspace{-0.5mm} V \approx 20.9$ cm/s, which for an original target mass of $M\approx1.9\times10^9$ kg results in a momentum enhancement factor of $\beta\sim15$, which is notably less than that achieved in the single penetrator case shown in Figure \ref{fig:100m100kgDamage}.}}
    \label{fig:100m5x100kgDamage}
\end{figure}

\textcolor{black}{As before, we also include a multiple penetrator case on the same 100 m diameter target. Figure \ref{fig:100m5x100kgDamage} shows a plot of the damage parameter at $t=10$ seconds after impact with an array of five penetrators in a similar ``+" configuration as for the earlier 50 m target, each with a measured mass of $\sim252.8$ for a total mass of $m_p\approx1264$ kg. At this time, we see an average damage parameter value of $\langle D\rangle=0.80$, which is approximately 5\% higher than the single penetrator case above. This is visualized by the increased extent and quantity of damaged regions in the plot of the damage parameter in Figure \ref{fig:100m5x100kgDamage}. At $t=10$ seconds after impact, we measure $\Delta\hspace{-0.5mm} V \approx 20.9$ cm/s, which for an original target mass of $M\approx1.9\times10^9$ kg results in a momentum enhancement factor of $\beta\sim15$. This is significantly lower than the single penetrator case presented earlier, which shows that distributing the mass across multiple penetrators yields two obvious effects: greater total material damage on average with increased number of off-axis penetrators, and significantly less momentum enhancement. The increase in damage for the multiple penetrator case can be expected from the interaction of multiple shockwaves within the target, which produces higher stress values on average due to significant constructive interference. The difference in momentum enhancement factor can be attributed to the multiple penetrator case producing ejecta at large angles away from the incident velocity vector, which contributes less to enhancing the momentum transfer than cases where the ejecta is more ``collimated" in the direction anti-parallel to the incident velocity vector.}

\subsection{\label{sec:altmodes}Alternative modes of operation}
While fragmentation of an incoming bolide appears to be effective in both short time scale terminal interdiction modes and in interdiction modes with long warning times, we also discuss two additional modes of operation which enable additional capability \cite{lubin_asteroid_2023,lubin_pi_2023}.

The first additional mode of operation involves the partial and asymmetric fragmentation of the bolide. This induces a force, and hence momentum transfer, on the remaining mass which has a component in the transverse direction (i.e. perpendicular to the initial bolide velocity vector). This hybrid disruption/deflection mode, which uses the ejected material to transfer momentum to the remaining mass, can operate in extended time scale interdiction modes with long warning times and also potentially for very large bolides in the $>100$ m class. Most kinetic impactor concepts intentionally avoid disrupting the asteroid, but not much work has been done on potential improved momentum transfer enhancement factors when partial disruption is achieved. This will be especially important for late warning time scenarios where there is no extended warning time to give the asteroid a gentle push. See our previous papers for a detailed discussion of the potentially large momentum transfer enhancement possible with this method \cite{lubin_asteroid_2023,lubin_pi_2023}.

If the closing speed of the bolide and penetrator is $>10$ km/s, then a significant portion of the kinetic energy of the penetrator is converted into hot vaporized bolide material within the first $10^4$ microseconds after impact (the rest is converted into shock waves within the bolide). This material is rapidly ejected through the entrance cavity caused by the penetrator impact, which induces a brief ``rocket mode" that yields a significant amount of imparted thrust to the remaining bolide. We have noticed this in a number of our simulations, and detailed modeling is forthcoming to get better estimates of the momentum impulse this provides. The thrust can be comparable to a heavy lift launcher for a short period of time ($\sim 1$ second) \cite{lubin_asteroid_2023,lubin_pi_2023}.

\section{\label{sec:future}\textcolor{black}{Future work}}
\textcolor{black}{A granite material was chosen for these precursor simulations in order to set an upper bound on the effect of material strength. Future simulations will explore alternative material models with less extreme strength distributions, such as basalt with a Weibull strength distribution. Additionally, while a porous crush model was implemented in these simulations, the porosity parameter was not varied. This is a critical parameter space that we are currently exploring, in addition to increasing both the impactor and target masses. Future work will extend the target size to 1 km diameter and penetrator mass up to 100 metric tons. We will also explore the effects of variations in the rubble pile internal structure, and we intend to explore the use of granular boulders in the place of perfect spheres, in addition to actual asteroid shape models as containers for our boulder distributions. This will allow us to simulate the DART impact as a direct comparison, and other hypothetical impacts with real asteroids, including those with extreme aspect ratios.} 

\textcolor{black}{The problem of targeting asteroids at 20 km/s closing speeds will also be explored in future work, including off-axis intercepts for cases where targeting is sub-optimal. However, there is precedence for $>10$ km/s targeting of $\sim1$ m size non-cooperative targets in the field of missile defense, such as the Patriot missile defense system's intercontinental ballistic missile (ICBM) intercept capability \cite{garwin_technical_1999}. In contrast, asteroids are passive targets and those of interest are generally $\geq10$ m in diameter, making them much more ``cooperative" targets than those successfully targeted by missile defense technologies.}

\section{\label{sec:conclusion}Conclusion}
\textcolor{black}{Using the LLNL hydrodynamics code ALE3D, we have shown via simulation that rubble pile asteroids in the 20 -- 100 meter-class can be effectively mitigated via disruptive or deflective means by way of 20 km/s impacts with 100 -- 1000 kg 10:1 aspect ratio cylindrical tungsten penetrators.} With such modest penetrator masses, the use of a SpaceX Falcon 9 (which has a payload capacity of $\sim2.5$ metric tons, and was used to launch the DART mission), or similarly capable launch vehicle, enables the mitigation of a wide variety of asteroid threats, likely well in excess of the largest 100 m bolide simulated in this study. The PI method enables operation in terminal interdiction modes where there is little warning time prior to impact, but can also operate in the same extended time scale interdiction modes as made possible by traditional deflection techniques. \textcolor{black}{The method is also practical and cost-effective since it relies solely on launch vehicles and penetrator materials readily available today, and thus presents itself as a logical and competitive option for planetary defense.}

\section{\label{sec:additional}Additional material}
A vast amount of further mathematical analyses, case studies, and simulation results pertaining to the acoustical and optical ground effects of the PI mitigation method has been collected on our group website: \url{www.deepspace.ucsb.edu/projects/pi-terminal-planetary-defense }

\section{Acknowledgments}
We gratefully acknowledge funding from NASA NIAC Phase I grant 80NSSC22K0764, NASA NIAC Phase II grant 80NSSC23K0966, and NASA California Space Grant NNX10AT93H, as well as from the Emmett and Gladys W. Fund. We would like to thank Dr. Richard Garwin, Dr. Glenn Sjoden, Dr. Irina Sagert, and Dr. Oleg Korobkin for their invaluable insight and expertise. We would also like to thank our large team of researchers and undergraduates, including Jeeya Khetia, Teagan Costa, Jasper Webb, Hannah Shabtian, Albert Ho, Alok Thakar, Alex Korn, Kellan Colburn, Daniel Bray, Kavish Kondap, and Jerry Zhang for their tireless work and enthusiasm to defend the planet.

\appendix

 \bibliographystyle{elsarticle-num-names} 


\bibliography{ALE3DAAPaper}




\end{document}